\newcommand{\paperOption}{twoside,comsoc}
\newcommand{\usehyperref}{}
\newcommand{\copyrightnotice}{}
\InputIfFileExists{option1}{%
 \typeout{Options overridden!}%
}{}

\documentclass[\paperOption]{IEEEtran}
\usepackage{amsmath,amsfonts,bm}
\usehyperref
\usepackage{mathenv}
\usepackage{graphicx}
\usepackage[noadjust]{cite}

\newtheorem{theorem}{Theorem}

\newtheorem{lemma}[theorem]{Lemma}
\newtheorem{proposition}[theorem]{Proposition}
\newtheorem{remark}[theorem]{Remark}
\newtheorem{definition}[theorem]{Definition}

\newcommand{\borel}{\mathcal{B}}
\newcommand{\ceil}[1]{\left\lceil{#1}\right\rceil}
\newcommand{\diff}{\mathrm{d}}

\newcommand{\eqdef}{:=}
\DeclareMathOperator*{\essinf}{ess\;inf}
\DeclareMathOperator{\expect}{\mathbb{E}}
\newcommand{\flip}[1]{\bar{#1}}
\newcommand{\floor}[1]{\left\lfloor{#1}\right\rfloor}
\newcommand{\nb}{\mathrm{e}}
\newcommand{\real}{\mathbb{R}}
 \newcommand{\nnreal}{\real_{\ge 0}}

\DeclareMathOperator{\mcr}{MCR}
\DeclareMathOperator{\nmcr}{NMCR}
\newcommand{\bernoulli}{\tilde{\mathrm{B}}}
\newcommand{\exponential}{\tilde{\mathrm{E}}}
\newcommand{\uniform}{\tilde{\mathrm{U}}}

\begin{document}

\title{A Maximin Optimal Online Power Control Policy for Energy Harvesting Communications}
\author{%
 Shengtian~Yang,~\IEEEmembership{Senior~Member,~IEEE,}
 and Jun~Chen,~\IEEEmembership{Senior~Member,~IEEE}%
 \thanks{%
  This work was supported in part by the National Natural Science Foundation of China under Grant 61571398 and in part by the Natural Sciences and Engineering Research Council (NSERC) of Canada under a Discovery Grant.
  A conference paper containing part of this paper is accepted for presentation at the 2020 IEEE International Conference on Communications.
  \copyrightnotice
 }
 \thanks{%
  S.~Yang is with the School of Information and Electronic Engineering, Zhejiang Gongshang University, Hangzhou 310018, China (e-mail: \mbox{yangst@codlab.net}).}
 \thanks{%
  J.~Chen is with the Department of Electrical and Computer Engineering, McMaster University, Hamilton, ON L8S 4K1, Canada (e-mail: \mbox{junchen@ece.mcmaster.ca}).}%
}

\maketitle

\begin{abstract}
	
A general theory of online power control for discrete-time battery limited energy harvesting communications is developed, which leads to, among other things, an explicit characterization of a maximin optimal policy. This policy	only requires the knowledge of the (effective) mean of the energy arrival process and maximizes the minimum asymptotic expected average reward 
(with the minimization taken over all energy arrival distributions  of a given (effective) mean).
Moreover, it is universally near optimal and has a strictly better worst-case performance as well as a strictly improved lower multiplicative factor in comparison with the fixed fraction policy proposed by Shaviv and \"{O}zg\"{u}r 
when the objective is to maximize the throughput over an additive white Gaussian noise channel.  The competitiveness of this maximin optimal policy is also demonstrated via numerical examples.


\end{abstract}


\IEEEpeerreviewmaketitle

\section{Introduction}

Recent advances in energy harvesting technologies have enabled the development of self-sustainable wireless communication systems that are powered by renewable energy sources in the environment. An important research topic of energy harvesting communications is to design power control policies that maximize throughput or other rewards under random energy availability (see, e.g., \cite{SMJG10,  OTYUY11, YU12, TY12, HZ12, OU12, BGD13, WL13, SK13, XZ14, RSV14, UYESZGH15, DFO15, shaviv_universally_2016, arafa_online_2018, ZC19, WZJC19}). Although offline power control is by far well investigated, our understanding of online power control remains quite limited.
This situation can be largely attributed to the  technical differences between these two  control problems. For offline power control, since the realization of the whole energy arrival process is known in advance, the underlying distribution is irrelevant as far as policy design is concerned and it enters the picture only in the evaluation of the expected reward, where
different realizations need to be weighted according to their respective probabilities. In contrast, for online power control, one has to take into account the distribution of the energy arrival process due to the uncertainty of future energy arrivals. Indeed, this fact can also be seen from the implicit characterization of the optimal online power control policy based on the Bellman equation, which invovles the energy arrival distribution in an essential way.
Due to its distribution-dependent nature, the Bellman equation is often very difficult to solve exactly. 
To the best of our knowledge,
the general analytical solution to the Bellman equation has only been found in the low battery-capacity regime where the greedy policy is shown to be optimal. Even in that case, the so-called low battery-capacity regime varies from one energy arrival distribution to another. More generally, for any nondegenerate reward function, there is no online power control policy that is universally optimal for all  energy arrival distributions, which should be contrasted with offline power control where universality comes for free in light of the aforementioned reason. It is worth mentioning that the requirement of precise knowledge of the energy arrival distribution not only complicates the characterization of the optimal online power control policy, but also, in a certain sense, diminishes the importance of such policy since the needed knowledge is typically not available in practice.

Fortunately, as demonstrated by Shaviv and \"{O}zg\"{u}r in their remarkable work \cite{shaviv_universally_2016}, it is possible to break the deadlock by weakening the notions of optimality and universality. Specifically, 
they proposed a fixed fraction policy, which only requires the knowledge of the (effective) mean of the energy arrival process, and established its universal near-optimality in terms of the achievable throughput over an additive white Gaussian noise (AWGN) channel (see also \cite{arafa_online_2018} for an extended version of this result for more general reward functions). At the heart of their argument is a worst-case performance analysis of the fixed fraction policy, which shows that among all energy arrival processes of the same (effective) mean, the Bernoulli process induces the minimum throughput for the fixed fraction policy; the aforementioned near-optimality result then follows directly from the fact that this minimum throughput is within both constant additive and multiplicative gaps from a simple universal upper bound. Their finding naturally raises the question of whether it is possible to find an online power control policy with improved worst-case performance as compared to the fixed fraction policy or, better still, a policy with the best worst-case performance. 
In this work, we provide an affirmative answer to this question by constructing an online power control policy that is maximin optimal in the following sense: this policy achieves the maximum asymptotic expected average reward for the Bernoulli energy arrival process of any (effective) mean while among all energy arrival processes of the same (effective) mean,  the Bernoulli process induces the minimum asymptotic expected average reward for this policy. To this end, two major obstacles need to be overcome.
First of all, the optimal online power control policy for the Bernoulli energy arrival process of a given (effective) mean is uniquely defined only for some discrete battery energy levels; however, under the maximin formulation, it is essential to extend the support of this policy to cover all possible battery energy levels, and a judicious construction is needed to ensure that the interpolated policy has desired properties and at the same time is amenable to analysis.
The second obstacle lies in the worst-case performance analysis of the interpolated policy. In contrast to the fixed fraction policy for which some basic convexity/concavity argument suffices due to its linearity, the interpolated policy requires more delicate reasoning for establishing Bernoulli arrivals as the least favorable form of energy arrivals. It will be seen that these two obstacles are intertwined, and we will address them by developing a maximin theory based on detailed investigations of some general families of online power control policies. From a mathematical perspective, our work can also be viewed as saddle-point analysis in a functional space. Note that even for finite-dimensional minimax/maximin games, one often relies on fixed-point theorems to prove the existence of saddle-point solutions.  It is thus somewhat surprising that the saddle-point solution of the specific functional game under consideration admits an explicit characterization. In this sense, our work is of inherent theoretical interest.

The rest of this paper is organized as follows.
In Sec.~\ref{mainResults}, we formulate the problem and introduce the main results of this paper.
A maximin theory of online power control for discrete-time battery limited energy harvesting communications is developed in Sec.~\ref{generalTheory}; this theory leads to an explicit characterization of a maximin optimal policy, which is shown to be universally near optimal and have a strictly better worst-case performance as well as a strictly improved lower multiplicative factor in comparison with the fixed fraction policy 
when the objective is to maximize the throughput over an additive white Gaussian noise channel. We conclude the paper in Sec.~\ref{conclusion}.
The proofs of Theorem~\ref{optimalPolicy} and most propositions, as well as some auxiliary results, are given in the appendices.

Throughout the paper, the base of the logarithm function is $\nb$.
The maximum and the minimum of $a$ and $b$ are denoted by $a\vee b$ and $a\wedge b$, respectively.
The Borel $\sigma$-field generated by the topology on a metric space $S$ is denoted by $\borel(S)$.
The $n$-fold product measure of a probability measure $Q$ is denoted by $Q^{\otimes n}$.
The $n$-fold composition of a function $f:A\to A$ for some subset $A$ of $\real$ is denoted by $f^{(n)}$ with the convention $f^{(0)}(x)=x$.
An empty sum and an empty product are defined to be $0$ and $1$, respectively.

\section{Problem Formulation and Main Results}\label{mainResults}

Consider a discrete-time energy harvesting communication system equipped with a battery of capacity $c>0$.
We denote by $x^\infty=(x_t)_{t=1}^\infty$ the amount of energy harvested at time $t=1,2,3,\ldots$.
An online power control policy $\pi^\infty=(\pi_t)_{t=1}^\infty$ is a family of mappings specifying the energy $u_t=\pi_t(x^t)$ consumed in time slot $t$ based on $x^t=(x_1,x_2,\ldots,x_t)$.
Let $b_{t^-}$ and $b_t$ denote the amounts of energy stored in the battery at the beginning of time slot $t$ before and after the arrival of energy $x_t$, respectively.
They satisfy
\begin{subequations}
\begin{eqnarray}
b_t
&= &(b_{t^-}+x_t)\wedge c,\\
b_{{(t+1)}^-}
&= &b_t-u_t.
\end{eqnarray}\label{batteryEnergyEquation}%
\end{subequations}
It is assumed that $b_{1^-}=0$.

A policy $\pi^\infty$ is said to be admissible if
\[
u_t\le b_t
\quad \text{for all $x^\infty\in\nnreal^\infty$ and all $t\ge 1$}.
\]
The collection of all admissible policies is denoted by $\Pi$.
For $\pi^\infty\in\Pi$, if $\pi_t$ depends on $x^t$ only through $b_t$ and is time invariant, we say $\pi^\infty$ is stationary and identify it by a mapping $\sigma: [0,c]\to [0,c]$ satisfying $\sigma(x)\le x$ for all $x\in [0,c]$ such that $\pi^\infty=(\sigma\circ b_t)_{t=1}^\infty$, where $b_t$ is understood as a function of $x^t$ by \eqref{batteryEnergyEquation}.
The set of all (admissible) stationary policies is denoted by $\Sigma$.
In the sequel, when we write a stationary policy $\sigma\in\Sigma$, it may be understood as a mapping $\sigma:[0,c]\to [0,c]$, a policy $(\sigma\circ b_t)_{t=1}^\infty$, or a partial policy $(\sigma\circ b_t)_{t=m}^n$, and so on, by the context.

The energy $u_t$ is consumed to perform some task in time slot $t$, from which a reward $r(u_t)$ is obtained.

\begin{definition}\label{reward.definition}
A reward function $r$ is a nondecreasing, Lipschitz, and concave function from $[0,+\infty)$ to $[0,+\infty)$ with $r(0)=0$.
\end{definition}

\begin{definition}\label{reward.regularity}
A reward function $r$ is said to be regular if it is strictly concave and differentiable and the function
\[
\kappa_s(x)
\eqdef r'^{-1}(sr'(x)),\quad x\in[\tau_s, +\infty)
\]
is convex for all $s>1$ (satisfying $sr'(+\infty)<r'(0)$, which is in fact unnecessary because $r'(+\infty)=0$ by Proposition~\ref{regularReward.property}), where
\[
\tau_s
\eqdef \kappa_{1/s}(0)
= r'^{-1}\left(\frac{r'(0)}{s}\right)
\in (0,+\infty).
\]
\end{definition}

One example of interest is the throughput over an AWGN channel.
In this case, the reward $r(u_t)$ is the information rate in time slot $t$ given by
\begin{equation}
r(u_t)
\eqdef \frac{1}{2}\ln(1+\gamma u_t)\;\text{(nats)}\label{awgnReward}
\end{equation}
with $\gamma$ being the channel coefficient.

Thus the $(m,n)$-horizon total reward of partial policy $\pi_m^n=(\pi_t)_{t=m}^n$ with respect to energy arrivals $x^n$ and the initial battery energy level $b_{m^-}$ is
\[
R_m^n(\pi_m^n,x^n,b_{m-})
\eqdef \sum_{t=m}^n r(u_t),
\]
where $m\le n$, $\pi\in\Pi$, and $u_t=\pi_t(x^t)\le b_t$, with $b_{t^-}$ and $b_t$ satisfying \eqref{batteryEnergyEquation}.
The corresponding $n$-horizon average reward is
\[
T_n(\pi^n,x^n)
\eqdef \frac{1}{n} R_1^n(\pi^n,x^n,0).
\]

Suppose now that the energy harvested at each time $t$ is a random variable $X_t$, and consequently the whole sequence $X^\infty=(X_t)_{t=1}^\infty$ of energy arrivals forms a random process.
Correspondingly, the energy variables $u_t$, $b_{t^-}$, and $b_t$ become the random variables $U_t$, $B_{t^-}$, and $B_t$, respectively.
The asymptotic expected average reward of policy $\pi^\infty$ with respect to energy arrivals $X^\infty$ is defined as
\[
\mathcal{T}(\pi^\infty,X^\infty)
\eqdef \liminf_{n\to\infty} \expect T_n(\pi^n,X^n).
\]
Since it depends only on the (probability) distribution of $X^\infty$, we can also write $P_{X^\infty}$ in place of $X^\infty$.
For example, we may write $\mathcal{T}(\pi^\infty,Q^{\otimes\infty})$, where $Q^{\otimes\infty}$ denotes the distribution of an i.i.d.\ process with marginal distribution $Q$.

We are interested in characterizing online power control policies that maximize the asymptotic expected average reward in the worst case of a given family of energy arrival distributions.
To this end, we introduce a maximin formulation.

\begin{definition}
The mean-to-capacity ratio (MCR) of a probability measure $Q$ on $(\nnreal,\borel(\nnreal))$ is defined by
\[
\mcr(Q)
\eqdef \frac{\mu_c(Q)}{c},
\]
where $\mu_c(Q)\eqdef\int (x\wedge c) \diff Q$ is the (effective) mean of $Q$.
\end{definition}

\begin{definition}
Let $\mathcal{Q}_{c,p}$ consist of all probability measures $Q$ on $([0,c],\borel([0,c]))$ with $\mcr(Q)=p$, where $p\in(0,1)$.
An online power control policy $\hat{\pi}^\infty$ is said to be maximin optimal for $\mathcal{Q}_{c,p}$ if
\[
\inf_{Q\in\mathcal{Q}_{c,p}} \mathcal{T}(\hat{\pi}^\infty,Q^{\otimes\infty})
= \sup_{\pi^\infty\in\Pi} \inf_{Q\in\mathcal{Q}_{c,p}} \mathcal{T}(\pi^\infty,Q^{\otimes\infty}).
\]
\end{definition}

The main result of this paper is summarized as follows.

\begin{theorem}[Theorems~\ref{maximinOptimalPolicy}, \ref{awgnReward.policy} and Proposition~\ref{optimalPolicy.inverse.property}]
If the reward function $r$ is regular, then the stationary policy
\[
\omega(x)
= \eta^{-1}(x)
\]
is maximin optimal for $\mathcal{Q}_{c,p}$ and its associated least favorable distribution is Bernoulli (see \eqref{bernoulli}), where
\[
\eta(x)
\eqdef \sum_{i=1}^{M(x)} \kappa_{1/(1-p)^{i-1}}(x)
= \sum_{i=1}^{\tilde{M}(x)} \kappa_{1/(1-p)^{i-1}}(x),
\]
and
\[
M(x)
\eqdef \ceil{\frac{\ln(r'(x)/r'(0))}{\ln(1-p)}},
\]
\[
\tilde{M}(x)
\eqdef \floor{\frac{\ln(r'(x)/r'(0))}{\ln(1-p)}}+1.
\]
In particular, if $r$ is given by \eqref{awgnReward}, then 
\[
\omega_\textsc{awgn}(x)
= \frac{1}{\gamma}\left[\frac{p(\gamma x+\tilde{M})}{1-(1-p)^{\tilde{M}}}-1\right]
\]
is maximin optimal, where $\tilde{M}$ is the least integer satisfying
\[
[1+p(\gamma x+\tilde{M})] (1-p)^{\tilde{M}}
< 1.
\]
\end{theorem}

\section{A Maximin Theory of Online Power Control for Energy Harvesting Communications}\label{generalTheory}

In order to find the maximin optimal online power control policy, we adopt the following approach:
\begin{enumerate}
	\item Find a distribution $\hat{Q}$ that is the least favorable one in $\mathcal{Q}_{c,p}$ when a policy in some special subset $\mathcal{N}$ of $\Pi$ or $\Sigma$ is employed.
	
	\item Construct a policy that is optimal for $\hat{Q}$ and is an element of $\mathcal{N}$.
\end{enumerate}
The rationale underlying this approach is best explained by the following fact.

\begin{proposition}\label{maximinOptimalityCondition}
Let $\hat{\pi}^\infty\in\Pi$.
If there is a distribution $\hat{Q}\in\mathcal{Q}_{c,p}$ such that
\begin{eqnarray*}
\mathcal{T}(\hat{\pi}^\infty,\hat{Q}^{\otimes\infty})
&= &\max_{\pi^\infty\in\Pi} \mathcal{T}(\pi^\infty,\hat{Q}^{\otimes\infty})\\
&= &\min_{Q\in\mathcal{Q}_{c,p}} \mathcal{T}(\hat{\pi}^\infty,Q^{\otimes\infty}),
\end{eqnarray*}
then $\hat{\pi}^\infty$ is maximin optimal.
\end{proposition}

\begin{IEEEproof}
For any $\pi^\infty$,
\begin{eqnarray*}
\inf_{Q\in\mathcal{Q}_{c,p}} \mathcal{T}(\pi^\infty,Q^{\otimes\infty})
&\le &\mathcal{T}(\pi^\infty,\hat{Q}^{\otimes\infty})\\
&\le &\max_{\pi^\infty\in\Pi} \mathcal{T}(\pi^\infty,\hat{Q}^{\otimes\infty})\\
&= &\min_{Q\in\mathcal{Q}_{c,p}} \mathcal{T}(\hat{\pi}^\infty,Q^{\otimes\infty}).
\end{eqnarray*}
So $\hat{\pi}^\infty$ is maximin optimal.
\end{IEEEproof}

\subsection{Normal Stationary Policies and the Least Favorable Distribution}

In this subsection, we will study a special family $\mathcal{N}$ of policies called normal (stationary) policies.
We will show that, for any $\sigma\in\mathcal{N}$, the Bernoulli distribution is the least favorable one in $\mathcal{Q}_{c,p}$ as long as $p$ is not below a certain threshold depending on $\sigma$.

\begin{definition}
For each (stationary) policy $\sigma\in\Sigma$, let $\flip{\sigma}$ be its associated policy induced by the complement operation:
\[
\flip{\sigma}(x)
\eqdef x-\sigma(x).
\]
Note that $\flip{\flip{\sigma}}=\sigma$.
\end{definition}

Policy $\flip{\sigma}$ may be called a (stationary) reserve policy because it specifies the amount of energy reserved for future use.

\begin{definition}\label{normalPolicyDefinition}
A policy $\sigma\in\Sigma$ is said to be normal if it is nondecreasing and concave.
The set of all normal policies is denoted by $\mathcal{N}$.
\end{definition}

\begin{proposition}\label{normalPolicy.property}
A normal policy $\sigma\in\mathcal{N}$ satisfies:
\begin{enumerate}
\item
$\sigma(0)=\flip{\sigma}(0)=0$.

\item
$\flip{\sigma}$ is nondecreasing and convex on $[0,c]$.

\item
Both $\sigma$ and $\flip{\sigma}$ are Lipschitz on $[0,c]$.

\item
Both $\sigma$ and $\flip{\sigma}$ are differentiable at all but at most countable points of $(0,c)$, and $\sigma'$ and $\flip{\sigma}'$ are nonincreasing and nondecreasing, respectively, on their domains of definition.
Moreover, both $\sigma'(x)$ and $\flip{\sigma}'(x)$ are between $0$ and $1$ whenever they exist.
\end{enumerate}
\end{proposition}

The proof of Proposition~\ref{normalPolicy.property} is given in Appendix~\ref{proofs_of_general_theory_section}.

The next theorem shows that the Bernoulli distribution is the least favorable one for normal policies under certain mild conditions.

\begin{theorem}\label{worstCaseOfNormalPolicy}
For a normal policy $\sigma\in\mathcal{N}$, if
\begin{equation}
r'(\sigma(x))\ge (1-p) r'(\sigma(\flip{\sigma}(x)))
\label{nongreedyCondition}
\end{equation}
for almost every $x\in [0,c]$, then
\[
\expect_{X^n\sim\bernoulli_p^{\otimes n}} R_1^n(\sigma,X^n,x)
\le \expect_{X^n\sim Q^{\otimes n}} R_1^n(\sigma,X^n,x)
\]
for all $Q\in\mathcal{Q}_{c,p}$ and $x\in [0,c]$, where
\begin{equation}\label{bernoulli}
\bernoulli_p
\eqdef (1-p)\delta_0+p\delta_{c},
\end{equation}
and
\begin{equation}
\delta_x(A)
\eqdef \begin{cases}
1, &if $x\in A$,\\
0, &otherwise.
\end{cases}\label{degenerate_probability_measure}
\end{equation}
\end{theorem}

\begin{IEEEproof}
Let $X^n$ and $\hat{X}^n$ be two random sequences of energy arrivals such that $P_{X^n}=Q^{\otimes n}$ and $P_{\hat{X}^n}=\bernoulli_p^{\otimes n}$, respectively.
Let
\[
f_t(x)
= \expect R_t^n(\sigma,X_t^n,x)
\]
and
\[
g_t(x)
= \expect R_t^n(\sigma,\hat{X}_t^{n},x),
\]
where $X_t^n=(X_i)_{i=t}^n$, $\hat{X}_t^n=(\hat{X}_i)_{i=t}^n$, $x\in [0,c]$, and $t=1,2,\ldots,n$.

Note that $\sigma$ and $r$ are both nondecreasing, concave, and Lipschitz (Definitions~\ref{reward.definition} and \ref{normalPolicyDefinition} and Proposition~\ref{normalPolicy.property}).
Let
\[
\chi(x,y)
= (x+y)\wedge c,
\]
where $x,y\ge 0$.
It is clear that $\chi$ is concave in $x$ for fixed $y$.
Recall that, for any concave functions $f$ and $g$, $f(g(x))$ is concave if $f$ is nondecreasing (\cite[p.~84]{boyd_convex_2004}). So a function such as $r(\sigma(\chi(x,y)))$ is concave in $x$ for fixed $y$.

Now we will show that $g_t(x)\le f_t(x)$ for all $1\le t\le n$.
Note that
\begin{eqnarray*}[rclqTl]
f_n(x)
&= &\expect r(\sigma(\chi(X_n,x)))\\
&\ge &\expect r(\sigma(\chi(\hat{X}_n,x))) &(\cite[Lemma~2]{shaviv_universally_2016})\\
&= &g_n(x)\\
&= &(1-p)r(\sigma(x))+pr(\sigma(c)),
\end{eqnarray*}
and
\[
g_n'(x)
= (1-p)r'(\sigma(x))\sigma'(x) \quad\text{a.e.};
\]
moreover,
\begin{eqnarray*}[rclqTl]
g_n'(\flip{\sigma}(x))
&= &(1-p)r'(\sigma(\flip{\sigma}(x)))\sigma'(\flip{\sigma}(x))\\
&\le &r'(\sigma(x))\sigma'(\flip{\sigma}(x)) &(Eq.~\eqref{nongreedyCondition})\\
&\le &r'(\sigma(x)) \quad\text{a.e.} &(Proposition~\ref{normalPolicy.property}),
\end{eqnarray*}
and $g_n(x)$ is nondecreasing, Lipschitz, and concave on $[0,c]$.

We proceed by induction on $t$ in the reverse order.
Suppose that
\begin{subequations}
\begin{eqnarray}
f_t(x)
&\ge &g_t(x),\label{worstCaseInduction.1}\\
g_t'(\flip{\sigma}(x))
&\le &r'(\sigma(x)) \quad\text{a.e.},\label{worstCaseInduction.2}
\end{eqnarray}
\end{subequations}
and $g_t(x)$ is nondecreasing, Lipschitz, and concave on $[0,c]$.
It follows that
\begin{eqnarray*}[rclqTl]
f_{t-1}(x)
&\ge &\expect h(\chi(X_{t-1},x)) &(Eq.~\eqref{worstCaseInduction.1})\\
&\ge &\expect h(\chi(\hat{X}_{t-1},x)) &(\cite[Lemma~2]{shaviv_universally_2016})\\
&= &g_{t-1}(x)\\
&= &(1-p)h(x)+ph(c)
\end{eqnarray*}
and
\[
g_{t-1}'(x)
\le (1-p)r'(\sigma(x)) \quad\text{a.e.}\quad\text{(Lemma~\ref{concavityIteration} with \eqref{worstCaseInduction.2})},
\]
which, together with Eq.~\eqref{nongreedyCondition}, implies
\[
g_{t-1}'(\flip{\sigma}(x))
\le r'(\sigma(x)) \quad\text{a.e.},
\]
where
\[
h(x)
= r(\sigma(x)) + g_t(\flip{\sigma}(x))
\]
is nondecreasing, Lipschitz, and concave on $[0,c]$ (Lemma~\ref{concavityIteration} with \eqref{worstCaseInduction.2}), and so is $g_{t-1}(x)$.
Therefore, $g_t(x)\le f_t(x)$ for all $t$, and in particular for $t=1$.
\end{IEEEproof}

\begin{remark}
In essence, condition~\eqref{nongreedyCondition} compares the marginal utilities of two energy consumptions specified by policy $\sigma$: one in the current time slot and the other in the next time slot if there is no new energy arrival, assuming that the distribution of energy arrivals is Bernoulli.
The marginal utilities of these two energy consumptions are
\[
r'(\sigma(x))
\text{ and }
(1-p)r'(\sigma(\flip{\sigma}(x))),
\]
respectively.
When condition~\eqref{nongreedyCondition} is met, $\sigma$ can be considered, in a certain sense,  non-greedy for $\mathcal{Q}_{c,p}$.
\end{remark}

Motivated by this observation, we introduce the following definitions.

\begin{definition}
A universal stationary policy $\sigma$ is a mapping from $\nnreal$ to $\nnreal$ satisfying $\sigma(x)\le x$.
The set of all universal stationary policy is denoted by $\Sigma_\infty$.
A universal stationary policy $\sigma\in\Sigma_\infty$ is said to be normal if it is nondecreasing and concave.
The set of all universal normal (stationary) policies is denoted by $\mathcal{N}_\infty$.
\end{definition}

Note that any universal stationary policy $\sigma$ can be regarded as a stationary policy in $\Sigma$ by considering its restriction on $[0,c]$.

\begin{definition}
The greed index $\iota_c(\sigma)$ of a stationary policy $\sigma\in\Sigma$ is defined by
\[
\iota_c(\sigma)
\eqdef 1-\essinf_{0\le x\le c} \frac{r'(\sigma(x))}{r'(\sigma(\flip{\sigma}(x)))}.
\]
The universal greed index $\iota(\sigma)$ of a universal stationary policy $\sigma\in\Sigma_\infty$ is defined by
\[
\iota(\sigma)
\eqdef 1-\essinf_{x\ge 0} \frac{r'(\sigma(x))}{r'(\sigma(\flip{\sigma}(x)))}.
\]
\end{definition}

\begin{remark}
Theorem~\ref{worstCaseOfNormalPolicy} only provides a sufficient condition, so it does not cover all possible policies for which the least favorable distribution is Bernoulli (e.g., the greedy policy $\sigma(x)=x$ for sufficiently large $c$ but fixed $p$).
However, since condition~\eqref{nongreedyCondition} coincides in part with the optimality condition for the Bernoulli distribution (see \eqref{infiniteKKT} with $\alpha_1=\sigma(x)$ and $\alpha_2=\sigma(\flip{\sigma}(x))$), any policy completely violating \eqref{nongreedyCondition} cannot be an optimal policy for the Bernoulli distribution, and hence is not maximin optimal even if its least favorable distribution is Bernoulli.
\end{remark}

We end this subsection with some properties of the greed index and the universal greed index. Their proofs are simple and hence left to the reader.

\begin{proposition}\label{greed_index.property}
Let $\sigma\in\Sigma$.
\begin{enumerate}
\item
$\iota_c(\sigma)\in [0,1]$.

\item
If $\iota_c(\sigma)\le p$, then $\sigma$ satisfies \eqref{nongreedyCondition}.
\end{enumerate}
\end{proposition}

\begin{proposition}\label{universalGreedIndex.property}
Let $\sigma\in\Sigma_\infty$.
\begin{enumerate}
\item
$\iota(\sigma)\in [0,1]$.

\item
$\iota_c(\sigma)$ is nondecreasing in $c$, and $\lim_{c\to +\infty} \iota_c(\sigma) = \iota(\sigma)$.

\item
If $\iota(\sigma)\le p$, then $\sigma$ satisfies \eqref{nongreedyCondition}.
\end{enumerate}
\end{proposition}

By Proposition~\ref{greed_index.property}, Theorem~\ref{worstCaseOfNormalPolicy} can be restated as follows:
If a normal policy $\sigma$ satisfies $\iota_c(\sigma)\le p$, then its least favorable distribution in $\mathcal{Q}_{c,p}$ is Bernoulli.

\subsection{An Optimal Policy for Bernoulli Energy Arrivals}

In this subsection we will construct an optimal policy for Bernoulli energy arrivals that is normal and satisfies $\iota_c(\sigma)\le p$, and consequently is maximin optimal.
In order to achieve this goal, the reward function is required to be regular (Definition~\ref{reward.regularity}).
This property further implies the following fact.

\begin{proposition}\label{regularReward.property}
A regular reward function $r$ is strictly increasing and continuously differentiable.
Its derivative $r'$ is strictly decreasing and satisfies $r'(+\infty)=0$.
\end{proposition}

\begin{IEEEproof}[Sketch of Proof]
Use \cite[Th.~1.5]{gruber_convex_2007} and Proposition~\ref{convexFunction.unbounded}.
\end{IEEEproof}

From now on, we will assume that $r$ is regular.
Under this assumption, we can construct an explicit optimal stationary policy for Bernouli energy arrivals.

\begin{definition}\label{bernoulli_throughput_notation}
For any universal stationary policy $\sigma$ parametrized by $p$, the asymptotic expected average reward $\mathcal{T}(\sigma,\bernoulli_p^{\otimes\infty})$ of $\sigma$ with respect to the Bernoulli energy arrival distribution $\bernoulli_p$ is a function of capacity $c$ and MCR $p$, which is denoted by $T_{\sigma}(c,p)$, or more succinctly, $T_\sigma(c)$, when $p$ is fixed and is clear from the context.
\end{definition}

\begin{theorem}[cf.\ {\cite[Th.~1]{shaviv_universally_2016} and \cite[Sec.~II-A]{arafa_online_2018}}]\label{optimalPolicy}
A stationary policy $\omega$ is optimal for i.i.d. energy arrivals with the Bernoulli distribution $\bernoulli_p$ iff it satisfies
\begin{equation}
x
= \sum_{i=1}^{M(\omega(x))} \kappa_{1/(1-p)^{i-1}}(\omega(x))\label{optimalPolicy.sum}
\end{equation}
for all
\begin{equation}
x\in C
\eqdef \left\{\flip{\omega}^{(i)}(c): 0\le i\le M(\omega(c))\right\}\label{ergodicSet},
\end{equation}
where
\begin{equation}
M(y)
\eqdef \ceil{\frac{\ln(r'(y)/r'(0))}{\ln(1-p)}}.\label{optimalPolicy.M}
\end{equation}
\end{theorem}

The proof of Theorem~\ref{optimalPolicy} is presented in Appendix~\ref{proofs_of_general_theory_section}.
From the proof of Theorem~\ref{optimalPolicy}, we can see that the value of $\omega(x)$ for $x\notin C$ has no impact on the asymptotic expected average reward for Bernoulli arrivals.
However, this is not necessarily the case for other energy arrival distributions. To construct a universal stationary policy with maximin optimal performance, we consider the natural extension of \eqref{optimalPolicy.sum} from $C$ to $\nnreal$. The resulting policy $\omega$ is analyzed with the aid of the following functions.

\begin{definition}
The extension $\bar{\kappa}_s$ of $\kappa_s$ is defined by
\[
\bar{\kappa}_s(x)
\eqdef \kappa_s(x\vee \tau_s)
= \begin{cases}
\kappa_s(x), &$x\ge\tau_s$,\\
0, &$0\le x<\tau_s$,
\end{cases}
\]
where $s>1$.
\end{definition}

\begin{definition}
Let
\[
\eta_s(x)
\eqdef \sum_{i=1}^\infty \bar{\kappa}_s^{(i-1)}(x),
\]
where $\bar{\kappa}_s^{(i)}$ denotes the $i$-fold composition of $\bar{\kappa}_s$ with the convention $\bar{\kappa}_s^{(0)}(x)=x$.
\end{definition}

The following propositions summarize some important properties of $\bar{\kappa}_s$ and $\eta_s$.

\begin{proposition}\label{conjugate.property}
The function $\bar{\kappa}_s$ has the following properties:
\begin{enumerate}
\item
$0\le \bar{\kappa}_s(x)<x$ for $x>0$.

\item
$\bar{\kappa}_s$ is continuous, nondecreasing, and convex.

\item
$\bar{\kappa}_s^{(i)}(x) = \kappa_{s^{i}}(x\vee \tau_s^{(i)})$, where $i\ge 0$ and $\tau_s^{(i)}\eqdef\kappa_{s^{-i}}(0)$.

\item
The least nonnegative integer $i$ such that $\bar{\kappa}_s^{(i)}(x)=0$ is
\[
M_s(x)
\eqdef \ceil{\frac{\ln(r'(0)/r'(x))}{\ln s}},
\]
which is a generalization of \eqref{optimalPolicy.M} (the latter  corresponds to the special case $s=1/(1-p)$).

\end{enumerate}
\end{proposition}

\begin{proposition}\label{optimalPolicy.inverse.property}
The function $\eta_s$ is continuous, strictly increasing, and convex, and $\eta_s(x)=\sum_{i=1}^{N} \bar{\kappa}_s^{(i-1)}(x)$ for all $N\ge M_s(x)$.
In particular,
\[
\eta_s(x)
= \sum_{i=1}^{M_s(x)} \kappa_{s^{i-1}}(x)
= \sum_{i=1}^{\tilde{M}_s(x)} \kappa_{s^{i-1}}(x),
\]
where
\[
\tilde{M}_s(x)
\eqdef \floor{\frac{\ln(r'(0)/r'(x))}{\ln s}}+1.
\]
\end{proposition}

The proofs of Propositions~\ref{conjugate.property} and \ref{optimalPolicy.inverse.property} are given in Appendix~\ref{proofs_of_general_theory_section}.
A straightforward consequence of these properties is the next theorem, which shows that $\omega$ is normal and $\iota(\omega)=p$.

\begin{theorem}\label{optimalPolicy.property}
The stationary policy $\omega(x)=\eta_{1/(1-p)}^{-1}(x)$ has the following properties:
\begin{enumerate}
\item
Policy $\omega$ is strictly increasing, concave, and consequently  normal.

\item\label{optimalPolicy.property.2}
\begin{eqnarray*}
\multicolumn{3}{l}{\omega(\flip{\omega}^{(i-1)}(x))
= \bar{\kappa}_{1/(1-p)}^{(i-1)}(\omega(x))}\\
\quad &= &\begin{cases}
\kappa_{1/(1-p)^{i-1}}(\omega(x)), &if $i\le M(\omega(x))$,\\
0, &otherwise.
\end{cases}
\end{eqnarray*}

\item
$\iota(\omega)=p$.
\end{enumerate}
\end{theorem}

\begin{IEEEproof}
1) It is clear that $x=\eta_{1/(1-p)}(\omega(x))$ (Theorem~\ref{optimalPolicy} and Proposition~\ref{optimalPolicy.inverse.property}), which implies $\omega(x)=\eta_{1/(1-p)}^{-1}(x)$ due to the invertibility of $\eta_{1/(1-p)}$  (Proposition~\ref{optimalPolicy.inverse.property}).
Moreover, $\omega$ is strictly increasing and concave (Propositions~\ref{optimalPolicy.inverse.property} and \ref{convexityOfInverse}).
Therefore, $\omega$ is normal.

2)
Since $x=\eta_{1/(1-p)}(\omega(x))$,
\begin{eqnarray*}
\flip{\omega}(x)
&= &\sum_{i=2}^\infty \bar{\kappa}_{1/(1-p)}^{(i-1)}(\omega(x))\\
&= &\sum_{i=1}^\infty \bar{\kappa}_{1/(1-p)}^{(i-1)}(\bar{\kappa}_{1/(1-p)}(\omega(x)))\\
&= &\eta_{1/(1-p)}(\bar{\kappa}_{1/(1-p)}(\omega(x)))\\
&= &\omega^{-1}(\bar{\kappa}_{1/(1-p)}(\omega(x))),
\end{eqnarray*}
which implies that $\omega(\flip{\omega}(x))=\bar{\kappa}_{1/(1-p)}(\omega(x))$.
Repeatedly applying this identity, we have $\omega(\flip{\omega}^{(i-1)}(x))=\bar{\kappa}_{1/(1-p)}^{(i-1)}(\omega(x))$, which is zero if $i>M_{1/(1-p)}(\omega(x))$ (Proposition~\ref{conjugate.property}).

3) It is clear that
\[
r'(\omega(\flip{\omega}(x)))
= \frac{r'(\omega(x)\vee\tau_{1/(1-p)})}{1-p}
\quad\text{(Property~\eqref{optimalPolicy.property.2})}.
\]
We have
\begin{eqnarray*}
\inf_{x\ge 0} \frac{r'(\omega(x))}{r'(\omega(\flip{\omega}(x)))}
&= &(1-p) \inf_{x\ge 0} \frac{r'(\omega(x))}{r'(\omega(x)\vee\tau_{1/(1-p)})}\\
&= &1-p,
\end{eqnarray*}
and therefore $\iota(\omega)=p$.
\end{IEEEproof}

From Theorems~\ref{worstCaseOfNormalPolicy}, \ref{optimalPolicy}, and \ref{optimalPolicy.property} and Proposition~\ref{maximinOptimalityCondition}, it then follows that $\omega$ is maximin optimal.

\begin{theorem}\label{maximinOptimalPolicy}
Suppose that $r$ is regular.
The stationary policy $\omega(x)=\eta_{1/(1-p)}^{-1}(x)$ is maximin optimal for $\mathcal{Q}_{c,p}$ and
\[
\inf_{Q\in\mathcal{Q}_{c,p}} \mathcal{T}(\omega,Q^{\otimes\infty})
= T_\omega(c)
\eqdef \mathcal{T}(\omega,\bernoulli_p^{\otimes\infty}),
\]
where $\bernoulli_p$ is the Bernoulli distribution defined by \eqref{bernoulli}.
\end{theorem}

In particular, for the special reward function given by \eqref{awgnReward}, we have the following maximin optimal policy.

\begin{theorem}[cf.\ {\cite[Th.~1]{shaviv_universally_2016}}]\label{awgnReward.policy}
Suppose that $r$ is given by \eqref{awgnReward}.
The policy 
\begin{equation}
\omega_\textsc{awgn}(x)
= \frac{1}{\gamma}\left[\frac{p(\gamma x+\tilde{M})}{1-(1-p)^{\tilde{M}}}-1\right] \label{awgnReward.policy.optimal}
\end{equation}
is maximin optimal, 
where $\tilde{M}$ is the least integer satisfying
\[
[1+p(\gamma x+\tilde{M})] (1-p)^{\tilde{M}}
< 1.
\]
\end{theorem}

\begin{IEEEproof}
With no loss of generality, we assume $\gamma=1$.
Note that
\[
r'(x)
= \frac{1}{2(1+x)},
\]
and
\[
r'^{-1}(x)
= \frac{1}{2x}-1\quad\text{for $x\in \left(0,\frac{1}{2}\right)$}.
\]
We have
\[
\kappa_s(x)
= r'^{-1}(sr'(x))
= \frac{1+x}{s}-1\quad\text{for $x\in (\tau_s,+\infty)$},
\]
and consequently
\[
\bar{\kappa}_s(x)
= \begin{cases}
\displaystyle\frac{1+x}{s}-1, &$x\ge\tau_s$,\\[0.25em]
0, &$0\le x<\tau_s$,
\end{cases}
\]
where $\tau_s=s-1$.
It is easy to see that $r$ is regular.

In light of Theorem~\ref{maximinOptimalPolicy}, the online power control policy
\[
\omega_\textsc{awgn}(x)
= \eta_{1/(1-p)}^{-1}(x)
\]
is maximin optimal. Note that
\begin{eqnarray*}
\eta_{1/(1-p)}(x)
&= &\sum_{i=1}^{\tilde{M}(x)} \left[(1+x)(1-p)^{i-1}-1\right]\\
&= &(1+x)\frac{1-(1-p)^{\tilde{M}(x)}}{p}-\tilde{M}(x)
\end{eqnarray*}
with
\[
\tilde{M}(x)
= \floor{-\frac{\ln(1+x)}{\ln(1-p)}}+1
> -\frac{\ln(1+x)}{\ln(1-p)}.
\]
Thus 
\[
\omega_\textsc{awgn}(x)
= \frac{p(x+\tilde{M})}{1-(1-p)^{\tilde{M}}}-1,
\]
where $\tilde{M}$ is the least integer satisfying
\[
[1+p(x+\tilde{M})] (1-p)^{\tilde{M}}
< 1.
\]
By replacing $x$ and $\omega_\textsc{awgn}(x)$ with $\gamma x$ and $\gamma\omega_\textsc{awgn}(x)$, respectively, we get \eqref{awgnReward.policy.optimal} for a general $\gamma$.
\end{IEEEproof}


\begin{figure*}[htbp]
\centering
\includegraphics{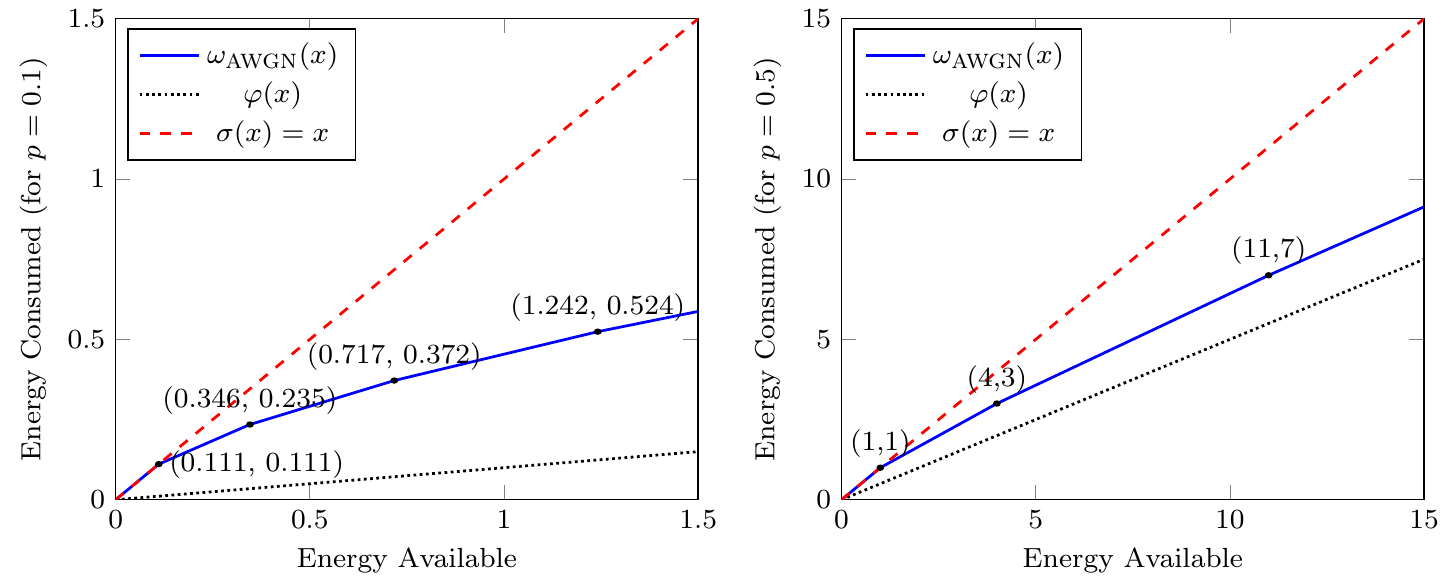}
\caption{Plots of the maximin opitmal policy $\omega_\textsc{awgn}(x)$ (for $\gamma=1$), the fixed fraction policy $\varphi(x)$, and the greedy policy $\sigma(x)=x$.}\label{policyGraph}
\end{figure*}

By Proposition~\ref{conjugate.property}, it is easy to see that $\omega_\textsc{awgn}$ is a piecewise linear function, with the endpoints of line segments given by
\begin{eqnarray}
E_k
&= &\frac{1}{\gamma} (\eta_{1/(1-p)}(\tau_{1/(1-p)}^{(k)}), \tau_{1/(1-p)}^{(k)})\nonumber\\
&= &\frac{1}{\gamma} \left(\frac{(1-p)^{-k}-1}{p}-k,(1-p)^{-k}-1\right)\label{endpoint}
\end{eqnarray}
for $k\ge 0$.
Policy $\omega_\textsc{awgn}$ is plotted in Fig.~\ref{policyGraph} for $\gamma=1$ and $p=0.1,0.5$.
For comparison, the fixed fraction policy
\begin{equation}
\varphi(x)
\eqdef px \label{fixed_fraction_policy}
\end{equation}
and the greedy policy $\sigma(x)=x$ are also plotted in Fig.~\ref{policyGraph}.
It is observed from \eqref{endpoint} and Fig.~\ref{policyGraph} that $\omega_\textsc{awgn}(x)$ coincides with $\sigma(x)$ when $x\leq p/(1-p)$.
It is also observed that $\omega_\textsc{awgn}(x)=\varphi(x)+\mathrm{O}(\ln x)$ as $x\to\infty$.

\subsection{Maximin Optimal Policy versus Fixed Fraction Policy}

In general, a maximin optimal policy does not necessarily perform well for all distributions in $\mathcal{Q}_{c,p}$.
But in the case where the reward function is given by \eqref{awgnReward}, there exists a certain performance guarantee as shown in the sequel.



\begin{definition}\label{lowerMultiplicativeFactor}
The  \emph{additive gap} $G(\pi^\infty, Q)$ and the 
 \emph{multiplicative factor} $F(\pi^\infty,Q)$ of $\pi^\infty\in\Pi$ for distribution $Q$ are defined respectively as
 \begin{eqnarray}
 &G(\pi^\infty, Q)
 \eqdef \sup_{\hat{\pi}^\infty} \mathcal{T}(\hat{\pi}^\infty,Q^{\otimes\infty}) - \mathcal{T}(\pi^\infty,Q^{\otimes\infty}),\nonumber\\ 
&F(\pi^\infty,Q)
\eqdef \frac{\mathcal{T}(\pi^\infty,Q^{\otimes\infty})}{\sup_{\hat{\pi}^\infty} \mathcal{T}(\hat{\pi}^\infty,Q^{\otimes\infty})}.\nonumber
\end{eqnarray}
When $\pi^\infty$ is parametrized by $(c,p)$ (and is written as $\pi_{c,p}^\infty$, or simply $\pi^\infty$ when there is no danger of confusion), the \emph{upper additive gap} and the \emph{lower multiplicative factor} of $\pi^\infty$ are defined respectively as
\begin{eqnarray}
&\overline{G}(\pi^\infty)
\eqdef \sup_{\substack{c>0\\ 0<p<1}} \overline{G}(\pi_{c,p}^\infty,c,p),\nonumber\\
&\underline{F}(\pi^\infty)
\eqdef \inf_{\substack{c>0\\ 0<p<1}} \underline{F}(\pi_{c,p}^\infty,c,p)\nonumber
\end{eqnarray}
with
\begin{eqnarray}
&\overline{G}(\pi_{c,p}^\infty,c,p)
\eqdef \sup_{Q\in\mathcal{Q}_{c,p}} G(\pi_{c,p}^\infty,Q),\nonumber\\
&\underline{F}(\pi_{c,p}^\infty,c,p)
\eqdef \inf_{Q\in\mathcal{Q}_{c,p}} F(\pi_{c,p}^\infty,Q).\nonumber
\end{eqnarray}
\end{definition}

According to (\cite[Th.~2]{shaviv_universally_2016}), the fixed fraction policy
$\varphi$ is universally near optimal for reward \eqref{awgnReward} in the sense that
\begin{subequations}\label{phi.performance_gap}
\begin{eqnarray}
&\overline{G}(\varphi)\leq\frac{1}{2},\\
&\underline{F}(\varphi)\geq\frac{1}{2}.\label{phi.multiplicative_factor}
\end{eqnarray}
\end{subequations}
This universal near optimality is established by considering the worst-case performance of $\varphi$ and invoking the fact that 
\(
\mathcal{T}(\hat{\pi}^\infty, Q^{\otimes\infty})
\le r(pc)
\)
for all $\hat{\pi}^\infty$ and $Q\in\mathcal{Q}_{c,p}$ (\cite[Prop.~2]{shaviv_universally_2016} and \cite[Lemma~1]{arafa_online_2018}).
Note that for both $\omega_\textsc{awgn}$ and $\varphi$, the least favorable distribution is Bernoulli (Theorem~\ref{worstCaseOfNormalPolicy} or \cite[Prop.~5]{shaviv_universally_2016}). 
Since $\omega_\textsc{awgn}$ is optimal for Bernoulli arrivals whereas $\varphi$ is suboptimal, it follows that  $\omega_\textsc{awgn}$ has a strictly better worst-case performance compared to $\varphi$ and consequently must be universally near optimal as well (in the sense of \eqref{phi.performance_gap} with $\varphi$ replaced by $\omega_\textsc{awgn}$).
The next result reveals that $\omega_\textsc{awgn}$ is actually superior to $\varphi$ in terms of the lower multiplicative factor.




\begin{theorem}\label{awgnReward.multiplicativeFactor}
For reward \eqref{awgnReward},
\begin{equation}
\underline{F}(\omega_\textsc{awgn})
\ge 1-\nb^{-1}
\approx 0.6321\label{multiplicativeFactor.optimal.awgn}
\end{equation}
while
\begin{equation}
\underline{F}(\varphi)
= \frac{1}{2}.\label{multiplicativeFactor.fp.awgn}
\end{equation}
\end{theorem}

\begin{IEEEproof}
Since the least favorable distribution for $\omega_\textsc{awgn}$
is Bernoulli and
\(
\mathcal{T}(\hat{\pi}^\infty, Q^{\otimes\infty})
\le r(pc)
\)
for all $\hat{\pi}^\infty$ and $Q\in\mathcal{Q}_{c,p}$ (\cite[Prop.~2]{shaviv_universally_2016} and \cite[Lemma~1]{arafa_online_2018}), we immediately have
\begin{eqnarray}
\underline{F}(\omega_\textsc{awgn},c,p)
&\ge &\frac{T_{\omega_\textsc{awgn}}(c,p)}{r(pc)}\nonumber\\
&= &\frac{\displaystyle\int_0^c \frac{\partial T_{\omega_\textsc{awgn}}(x,p)}{\partial x} \diff x}{r(pc)}\nonumber\\
&= &\frac{\displaystyle\int_0^c pr'(\omega_\textsc{awgn}(x)) \diff x}{r(pc)} \qquad\text{(Lemma~\ref{bernoulli.performance.derivative})}\nonumber\\
&\ge &\frac{\displaystyle\int_0^c pr'(px)\essinf_{y>0} \frac{r'(\omega_\textsc{awgn}(y)}{r'(py)} \diff x}{r(pc)}\nonumber\\
&= &\essinf_{y>0} \frac{r'(\omega_\textsc{awgn}(y))}{r'(py)}.\label{ratio_lower_bound}
\end{eqnarray}
With no loss of generality, we assume $\gamma=1$.
Then from Theorem~\ref{awgnReward.policy} with $\gamma=1$, it follows that
\begin{eqnarray*}
\frac{r'(\omega_\textsc{awgn}(x))}{r'(px)}
&= &\frac{(1+px)\left[1-(1-p)^{\tilde{M}}\right]}{p(x+\tilde{M})}\\
&= &\left[1-(1-p)^{\tilde{M}}\right]\left(1+\frac{1-p\tilde{M}}{p(x+\tilde{M})}\right),
\end{eqnarray*}
which is nonincreasing for $\tilde{M}\le 1/p$ and is increasing for $\tilde{M}>1/p$.
So its minimum is attained at
\begin{eqnarray*}[rcl]
x_0
&= &\sup\big\{x\colon [1+p(x+\tilde{M}_0)] (1-p)^{\tilde{M}_0}<1,\\
& &\qquad [1+p(x+\tilde{M}_0-1)] (1-p)^{\tilde{M}_0-1}\ge 1\big\}\\
&= &\frac{(1-p)^{-\tilde{M}_0}-1}{p}-\tilde{M}_0,
\end{eqnarray*}
where $\tilde{M}_0=\floor{1/p}$, and the minimum ratio is
\begin{eqnarray*}
F_0(p)
&= &\left[1-(1-p)^{\tilde{M}_0}\right]\left(1+\frac{1-p\tilde{M}_0}{(1-p)^{-\tilde{M}_0}-1}\right)\\
&= &1-(1-p)^{\tilde{M}_0}+(1-p\tilde{M}_0)(1-p)^{\tilde{M}_0}\\
&= &1-p\tilde{M}_0(1-p)^{\tilde{M}_0}.
\end{eqnarray*}
For $p\in [1/(1+n), 1/n)$ where $n\ge 1$,
\[
F_0(p)
= 1 - np(1-p)^n,
\]
and hence
\[
F_0'(p)
= n(1-p)^{n-1}[(n+1)p-1]
\ge 0,
\]
so that the minimum of $F_0(p)$ over $p\in [1/(1+n), 1/n)$ is
\[
F_0\left(\frac{1}{n+1}\right)
= 1 - \left(1-\frac{1}{n+1}\right)^{n+1}.
\]
Therefore,
\begin{eqnarray*}[rclqTl]
\underline{F}(\omega_\textsc{awgn})
&= &\inf_{0<p<1} \inf_{c>0} \underline{F}(\omega,c,p) &(Eq.~\eqref{ratio_lower_bound})\\
&\ge &\inf_{0<p<1} F_0(p)\\
&= &\inf_{n\ge 1} \left[1-\left(1-\frac{1}{n+1}\right)^{n+1}\right]
= 1-\nb^{-1}.
\end{eqnarray*}

Now let us evaluate $\underline{F}(\varphi)$.
In view of \eqref{phi.multiplicative_factor}, it suffices to show that $\underline{F}(\varphi)\le 1/2$.
For fixed $p$, by Lemmas~\ref{bernoulli.performance} and \ref{bernoulli.performance.derivative} and the dominated convergence theorem, we have
\begin{eqnarray*}
\lim_{c\to 0} F(\varphi,\bernoulli_{p})
&= &\lim_{c\to 0} \sum_{i=1}^\infty p(1-p)^{i-1} \frac{r(\varphi(\flip{\varphi}^{(i-1)}(c)))}{T_{\omega_\textsc{awgn}}(c)}\\
&= &\sum_{i=1}^\infty p(1-p)^{i-1} \lim_{c\to 0} \frac{r(\varphi(\flip{\varphi}^{(i-1)}(c)))}{T_{\omega_\textsc{awgn}}(c)}\\
&= &\sum_{i=1}^\infty p(1-p)^{i-1} \frac{p(1-p)^{i-1}r'(0)}{pr'(0)}
= \frac{1}{2-p},
\end{eqnarray*}
so that $\underline{F}(\varphi) \le \inf_{0<p<1} 1/(2-p) = 1/2$.
\end{IEEEproof}

\begin{figure*}[htbp]
\centering
\includegraphics{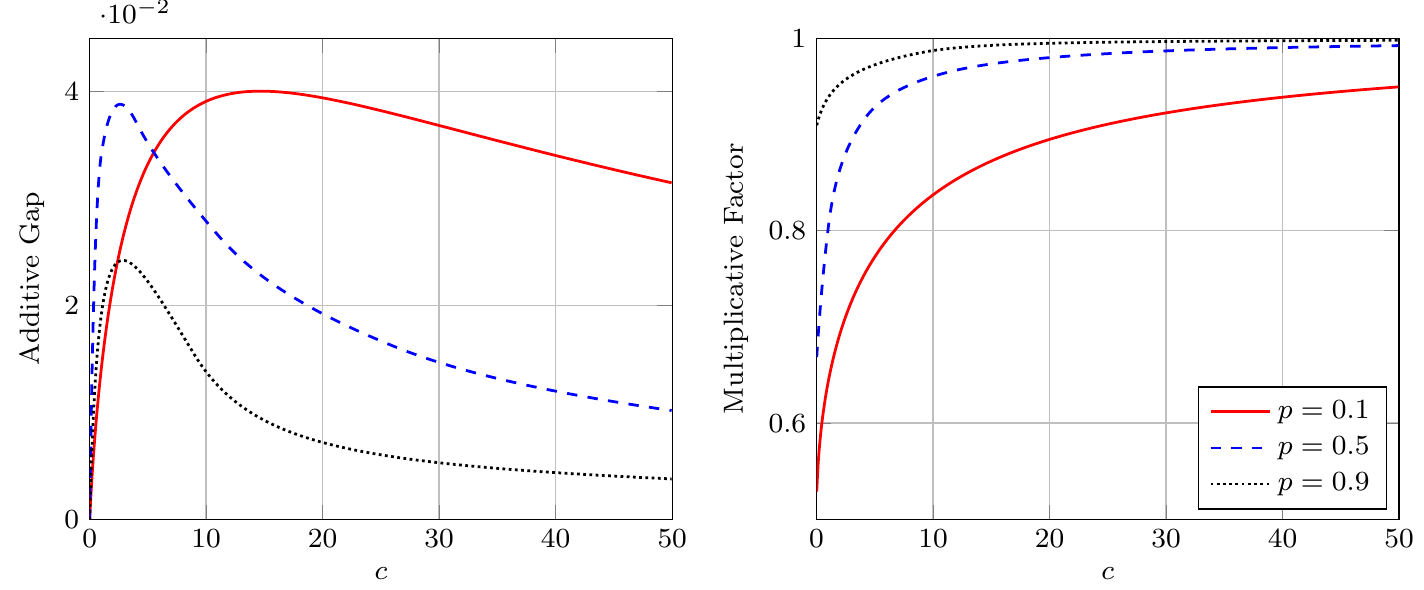}
\caption{The additive gaps and multiplicative factors of fixed fraction policies for reward \eqref{awgnReward} with $\gamma=1$ and $X_t\sim\bernoulli_p$ with $p=0.1, 0.5, 0.9$, respectively.}\label{first}
\end{figure*}

\begin{figure*}[htbp]
\centering
\includegraphics{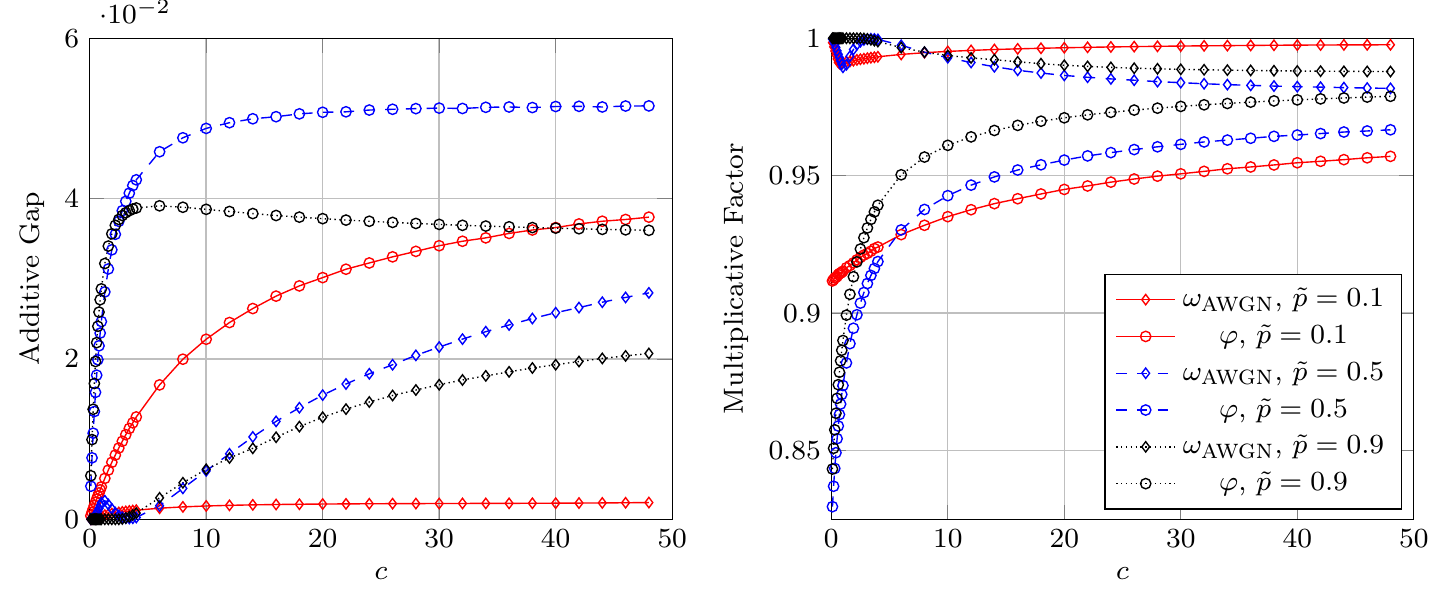}
\caption{The additive gaps and multiplicative factors of maximin optimal and fixed fraction policies for reward \eqref{awgnReward} with $\gamma=1$ and $X_t\sim\uniform_{2\tilde{p}c}$ with $\tilde{p}=0.1,0.5,0.9$ ($\mcr=0.1, 0.5$, and approximately $0.7222$), respectively.}
\end{figure*}

\begin{figure*}[htbp]
\centering
\includegraphics{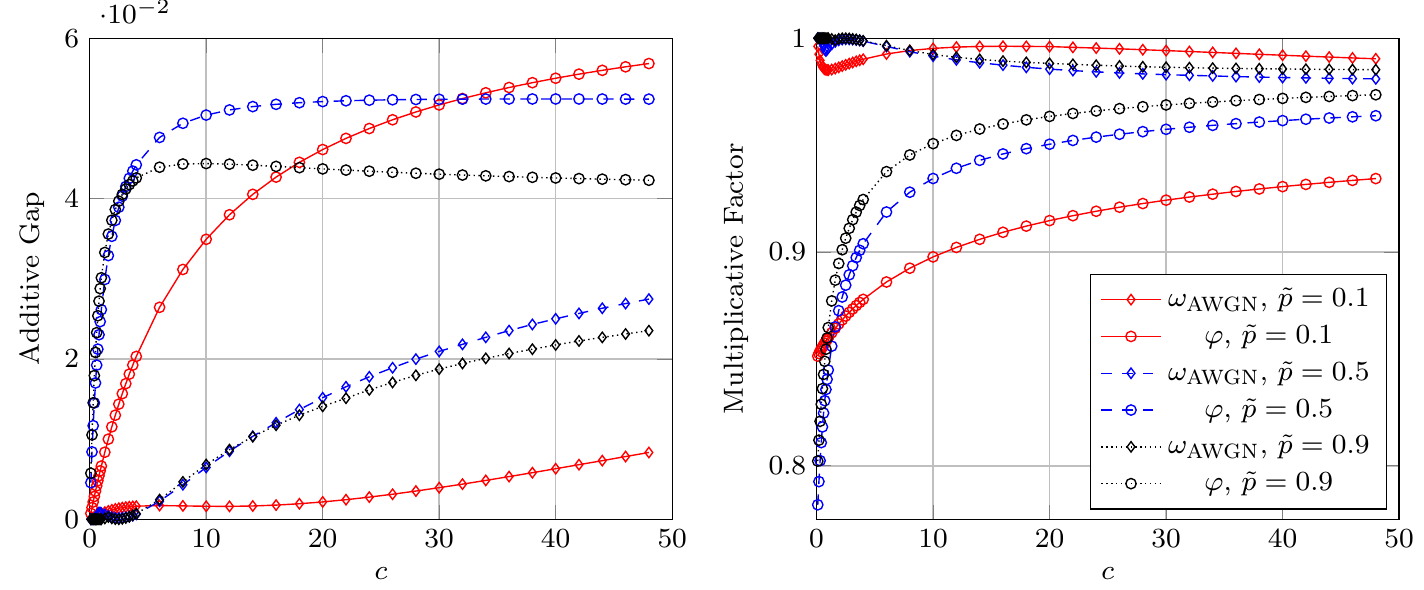}
\caption{The additive gaps and multiplicative factors of maximin optimal and fixed fraction policies for reward \eqref{awgnReward} with $\gamma=1$ and $X_t\sim\exponential_{(\tilde{p}c)^{-1}}$ with $\tilde{p}=0.1,0.5,0.9$ ($\mcr\approx 0.1000, 0.4323, 0.6037$), respectively.}\label{last}
\end{figure*}

Figs.~\ref{first}--\ref{last} illustrate the performance comparisons of the maximin optimal policy $\omega_\textsc{awgn}$ and the fixed fraction policy $\varphi$ when $Q$ is Bernoulli, $c$-limited uniform, or $c$-limited exponential, where the $c$-limited uniform and $c$-limited exponential distributions are given by
\begin{equation}
\uniform_b(A)
\eqdef \int_A \frac{1}{b} 1\{0\le x\le b\wedge c\} \diff x + \left(1-\frac{c}{b}\right)_+ \delta_c(A) \label{limited_uniform}
\end{equation}
with $(x)_+\eqdef x\vee 0$ and
\begin{equation}
\exponential_\lambda(A)
\eqdef \int_A \lambda \nb^{-\lambda x} 1\{0\le x\le c\} \diff x + \nb^{-\lambda c} \delta_c(A), \label{limited_exponential}
\end{equation}
respectively, where $\delta_c$ is defined by \eqref{degenerate_probability_measure}.
The comparison is performed by computing the additive gap
and the multiplicative factor (Definition~\ref{lowerMultiplicativeFactor}) of a policy $\pi^\infty$ for a distribution $Q$.
Note that $\omega_\textsc{awgn}$ is optimal in the Bernoulli case.
A modified value iteration algorithm based on \cite[Sec.~8.5]{puterman_markov_2005} is employed to compute the optimal performance as well as the performance of a given policy in non-Bernoulli cases.
It can be seen from the plots that $\omega_\textsc{awgn}$ consistently outperforms $\varphi$ and has a clear advantage in the low battery-capacity regime (i.e., when $c$ is small).
This shows that the dominance of $\omega_\textsc{awgn}$ over $\varphi$ is not restricted to the worst-case scenario.
Moreover, the performance of $\omega_\textsc{awgn}$ is very close (within 2\%) to the optimal in the non-Bernoulli cases under consideration, in particular for low MCRs.

\begin{figure*}[htbp]
\centering
\includegraphics{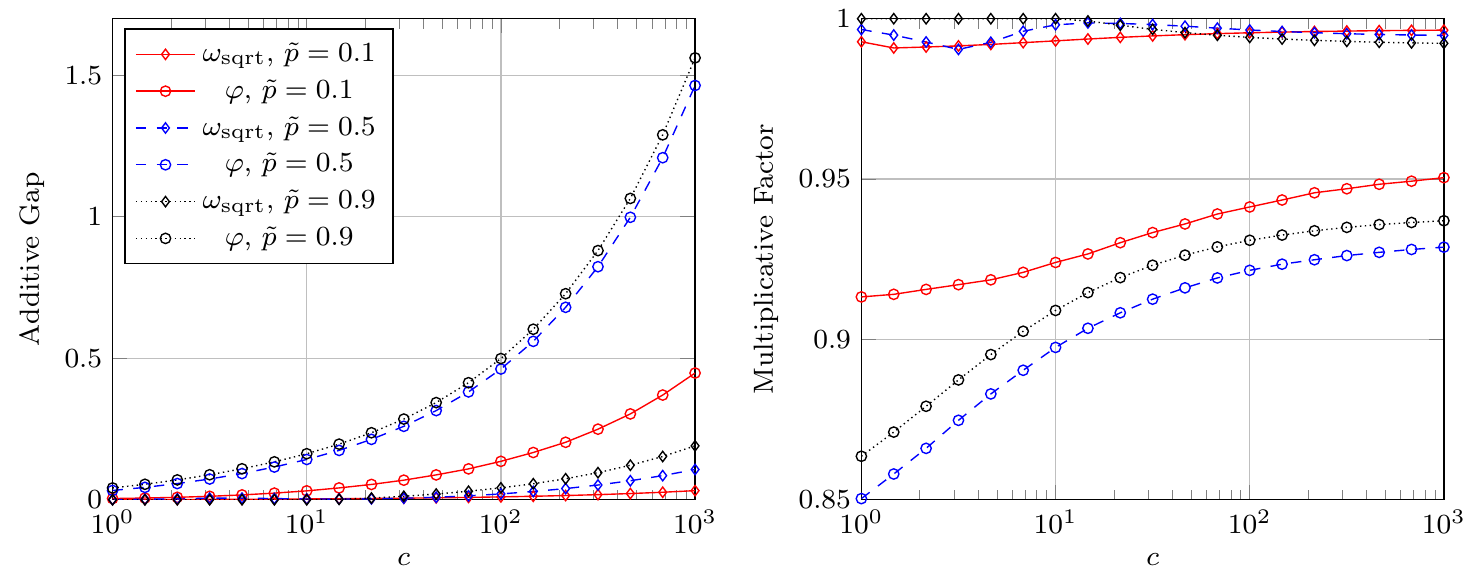}
\caption{The additive gaps and multiplicative factors of maximin optimal and fixed fraction policies for reward \eqref{sqrt_reward} and $X_t\sim\uniform_{2\tilde{p}c}$ with $\tilde{p}=0.1,0.5,0.9$ ($\mcr=0.1, 0.5$, and approximately $0.7222$), respectively.}\label{performance_for_sqrt_reward}
\end{figure*}

The advantage of the maximin optimal policy is potentially more evident for other reward functions. Indeed, for the square root reward
\begin{equation}
r(x)
\eqdef (1+x)^{1/2} - 1,\label{sqrt_reward}
\end{equation}
the maximin optimal policy, denoted by $\omega_\textrm{sqrt}$,  can outperform the fixed fraction policy by a wide margin, as is shown by Fig.~\ref{performance_for_sqrt_reward}.

\begin{remark}
In contrast to the fact that $\mcr(\bernoulli_p)=p$, the MCRs of $\uniform_b$ and $\exponential_\lambda$ depend on the battery capacity $c$ (in addition to their respective parameters $b$ and $\lambda$). To facilitate the characterization of this dependency, we define the nominal MCR (NMCR) of a battery-capacity-limited distribution to be the ratio of the mean of its original distribution to the battery capacity $c$. Note that the NMCRs of $\uniform_b$ and $\exponential_\lambda$ are
\[
\nmcr(\uniform_b)
= \frac{b}{2c}
\]
and
\[
\nmcr(\exponential_\lambda)
= \frac{1}{\lambda c},
\]
respectively.
Hence, if the NMCR is $\tilde{p}$, then
\[
\uniform_b=\uniform_{\tilde{p}\cdot 2c}
\]
and
\[ 
\exponential_\lambda=\exponential_{(\tilde{p}c)^{-1}},
\]
and their actual MCRs are
\begin{eqnarray*}
	\mcr(\uniform_b)
	&= &\begin{cases}
		\displaystyle \frac{b}{2c}, &$0\le b\le c$,\\
		\displaystyle 1-\frac{c}{2b}, &$b>c$,\\
	\end{cases}\\
	&= &\begin{cases}
		\tilde{p}, &$0\le \tilde{p}\le \frac{1}{2}$,\\
		\displaystyle 1-\frac{1}{4\tilde{p}}, &$\tilde{p}>\frac{1}{2}$,\\
	\end{cases}
\end{eqnarray*}
and
\[
\mcr(\exponential_\lambda)
= \frac{1}{\lambda c}(1 - \nb^{-\lambda c})
= \tilde{p}(1 - \nb^{-1/\tilde{p}}),
\]
respectively.

	
	

\end{remark}

\section{Conclusion}\label{conclusion}

We have constructed a maximin optimal online power control policy for discrete-time battery limited energy harvesting communications. This policy only requires the knowledge of the (effective) mean of the energy arrival process and achieves the best possible worst-case performance. 
It is of considerable interest to compare our new policy against the existing ones  in a systematic manner and quantify the performance gains. We have made some initial attempt along this direction and in particular showed that the new policy strictly outperforms the fixed fraction policy in terms of the lower multiplicative factor when the objective is to maximize the throughput over an additive white Gaussian noise channel.
It is also worthwhile to explore possible ways to simplify the new policy without essentially compromising its competitiveness or to enhance it for coping with the more realistic scenario where the reward function and the energy arrival distribution are not static. We intend to undertake some of these tasks in a follow-up work \cite{yang_ehpart2_2019}.

\appendices

\section{Proofs of Results in Sec.~\ref{generalTheory}}\label{proofs_of_general_theory_section}

\begin{IEEEproof}[Proof of Proposition~\ref{normalPolicy.property}]
1) $0\le \sigma(0)\le 0$.

2) It is clear that $\flip{\sigma}$ is convex.
For $y>x\ge 0$,
\begin{eqnarray*}
\flip{\sigma}(x)
&= &\flip{\sigma}\left(\frac{x}{y}y+\left(1-\frac{x}{y}\right)0\right)\\
&\le &\frac{x}{y}\flip{\sigma}(y)+\left(1-\frac{x}{y}\right)\flip{\sigma}(0)\le \flip{\sigma}(y).
\end{eqnarray*}

3) For $0\le x<y\le c$, we have
\begin{equation}
0
\le \frac{\sigma(y)-\sigma(x)}{y-x}
\le \frac{\sigma(y)-\sigma(0)}{y-0}
\le 1
\quad\text{(\cite[Lemma~1.1]{gruber_convex_2007})},\label{derivativeBound}
\end{equation}
which implies $|\sigma(y)-\sigma(x)|\le |y-x|$.
Therefore, $\sigma$ is Lipschitz on $[0,c]$, and so is $\flip{\sigma}$.

4) Use \cite[Th.~1.4]{gruber_convex_2007} and inequality \eqref{derivativeBound}.
\end{IEEEproof}

\begin{IEEEproof}[Proof of Theorem~\ref{optimalPolicy}]
We first prove the existence of a stationary policy that is optimal for the Bernoulli distribution.
Let $\mathfrak{f}\eqdef p\mathfrak{m}/c + \delta_c$ be a finite measure on $([0,c],\borel([0,c]))$, where $\mathfrak{m}$ denotes the Lebesgue measure on $[0,c]$.
Let $\epsilon = p$.
It is clear that any measurable set $A$ with $\mathfrak{f}(A)
\le \epsilon$ does not contain the point $c$, and consequently
\begin{eqnarray*}
P_{B_2\mid B_1}(A\mid b)
&\le &P_{B_2\mid B_1}([0,c)\mid b)\\
&= &\sum_{u\in [0,b]} P_{U_1\mid B_1}(u\mid b) P_{B_2\mid B_1,U_1}([0,c)\mid b,u)\\
&= &\sum_{u\in [0,b]} P_{U_1\mid B_1}(u\mid b) P\{b-u+X_1<c\}\\
&\le &\sum_{u\in [0,b]} P_{U_1\mid B_1}(u\mid b) (1-p)\\
&\le &1-p
= 1-\epsilon
\end{eqnarray*}
for all $b\in [0,c]$ and all (admissible) randomized stationary policies $P_{U_1\mid B_1}$.
This means that the so-called Doeblin condition is satisfied, and hence there exist a set $C\in\borel([0,c])$ with $\mathfrak{f}(C)>\epsilon$ and a stationary policy $\omega\in\Sigma$ such that for all $B_1=b\in C$, policy $\omega$ achieves the maximum asymptotic expected average reward and $P_{B_2\mid B_1}(C\mid b) = 1$ (\cite[Th.~2.2]{kurano_existence_1989}).
It is clear that $c\in C$, and in fact, by the property of Bernoulli distribution, $C$ is an invariant set consisting of the points
\[
\flip{\omega}^{(0)}(c)=c, \flip{\omega}^{(1)}(c), \ldots, \flip{\omega}^{(N-1)}(c), \flip{\omega}^{(N)}(c)=0
\]
for some integer $N$ to be determined later.
Note that for $B_{1^-}=0$, the distribution of $B_1$ is supported on $\{0,c\}\subseteq C$.\footnote{If the distribution of $B_1$ is supported on a set not contained in $C$, e.g., $[0,c]$, then the energy $B_t$ stored in the battery will undergo a transient stage, which however has negligible impact on the long-term expected average reward (see, e.g., \cite[Prop.~6]{shaviv_universally_2016} or \cite[Lemma~3.1]{kurano_existence_1989}).}
The asymptotic expected average reward of policy $\omega$ is
\[
T_\omega(c)
= \sum_{i=1}^\infty p(1-p)^{i-1} r(\omega(\flip{\omega}^{(i-1)}(c)))
\quad\text{(Lemma~\ref{bernoulli.performance})},
\]
or equivalently,
\[
T_\omega(c)
= \sum_{i=1}^\infty p(1-p)^{i-1} r(\alpha_i)
\]
with the constraint $\sum_{i=1}^\infty \alpha_i\le c$, where $\alpha_i=\omega(\flip{\omega}^{(i-1)}(c))\ge 0$.

In order to find $(\alpha_i)_{i=1}^\infty$, we need to solve the following infinite-dimensional optimization problem:
\begin{eqnarray*}[Tlql]
maximize &T((u_i)_{i=1}^\infty)\eqdef \sum_{i=1}^\infty p(1-p)^{i-1} r(u_i)\\
subject to &u_i\ge 0, \quad i=1,2,\ldots,\nonumber\\
&\sum_{i=1}^\infty u_i \le c,\nonumber
\end{eqnarray*}
where $c>0$.
It can be shown via an  argument similar to \cite[Appx.~C]{shaviv_universally_2016} that
\begin{subequations}\label{infiniteKKT}
\begin{equation}
r'(\alpha_i)
= \frac{\lambda_0}{p(1-p)^{i-1}}
\quad\text{for $1\le i\le M$},\label{infiniteKKT.1}
\end{equation}
\begin{equation}
\lambda_0
\ge p(1-p)^{M} r'(0),\label{infiniteKKT.2}
\end{equation}
\begin{equation}
\sum_{i=1}^M \alpha_i
= c.\label{infiniteKKT.3}
\end{equation}
\end{subequations}
Then, for $1\le i,j\le M$, we have
\[
\frac{r'(\alpha_i)}{r'(\alpha_j)}
= \frac{1}{(1-p)^{i-j}}
\]
and
\begin{eqnarray*}
(1-p)^M r'(0)
&\le &\frac{\lambda_0}{p}
= r'(\alpha_j)(1-p)^{j-1}\\
&= &r'(\alpha_M)(1-p)^{M-1}
< (1-p)^{M-1} r'(0).
\end{eqnarray*}
So
\begin{eqnarray*}
\sum_{i=1}^{M_j} \kappa_{1/(1-p)^{i-1}}(\alpha_j)
&= &\sum_{i=1}^{M_j} \alpha_{j+i-1}\\
&= &\sum_{i=j}^{M} \alpha_i
= \flip{\omega}^{(j-1)}(c)
\end{eqnarray*}
with
\[
M_j
= M-j+1
= \ceil{\frac{\ln(r'(\alpha_j)/r'(0))}{\ln(1-p)}}.
\]
Since $\alpha_j=\omega(\flip{\omega}^{(j-1)}(c))$, $\omega(x)$ satisfies \eqref{optimalPolicy.sum} for all $x\in C$ defined by \eqref{ergodicSet}, including $x=\flip{\omega}^{(M)}(c)=0$.
In view of the fact that $\kappa_s$ is strictly increasing (see also Proposition~\ref{optimalPolicy.inverse.property}),  Eq.~\eqref{optimalPolicy.sum} uniquely determines all $\alpha_i$, and we can conclude that $\omega$ is optimal iff it satisfies \eqref{optimalPolicy.sum}.
\end{IEEEproof}

\begin{IEEEproof}[Proof of Proposition~\ref{conjugate.property}]
1) It is clear that $\bar{\kappa}_s(x)\ge 0$.
Since $s>1$, it is also easy to see that $\bar{\kappa}_s(x)<r'^{-1}(r'(x))=x$ for $x>\tau_s$.

2) It is clear that $\bar{\kappa}_s(x)=f(g(x))$ with $f(x)=\kappa_s(x)$ and $g(x)=x\vee\tau_s$.
Since $r$ is regular, $f$ is continuous, strictly increasing, and convex on $[\tau_s,+\infty)$.
It is also clear that $g$ is continuous, nondecreasing, and convex on $[0,+\infty)$.
Therefore, $\bar{\kappa}_s$ is continuous, nondecreasing, and convex (\cite[p.~84]{boyd_convex_2004}).

3) It is clear that $\bar{\kappa}_s^{(i)}(x) = \kappa_{s^{i}}(x\vee \tau_s^{(i)})$ for $i=0,1$.
Suppose the identity is true for $i=k$.
Then
\begin{eqnarray*}
\bar{\kappa}_s^{(k+1)}(x)
&= &\kappa_s(\bar{\kappa}_s^{(k)}(x)\vee\tau_s)\\
&= &\kappa_s(\kappa_{s^{k}}(x\vee\tau_s^{(k)})\vee\tau_s)\\
&= &\kappa_s(\kappa_{s^{k}}(x\vee\tau_s^{(k)}\vee\kappa_{s^{-k}}(\tau_s)))\\
&= &\kappa_s(\kappa_{s^{k}}(x\vee\tau_s^{(k)}\vee\tau_s^{(k+1)}))\\
&= &\kappa_s(\kappa_{s^{k}}(x\vee\tau_s^{(k+1)}))
= \kappa_{s^{k+1}}(x\vee\tau_s^{(k+1)}).
\end{eqnarray*}
Therefore by induction, $\bar{\kappa}_s^{(i)}(x) = \kappa_{s^{i}}(x\vee\tau_s^{(i)})$ for all $i\ge 0$.

4) The least nonnegative integer $i$ such that $\bar{\kappa}_s^{(i)}(x)=0$ is exactly the least nonnegative integer $M$ satisfying
\[
x
\le \tau_s^{(M)}
= \kappa_{s^{-M}}(0),
\]
or equivalently,
\[
s^Mr'(x)
\ge r'(0).
\]
In other words,
\[
M = \ceil{\frac{\ln(r'(0)/r'(x))}{\ln s}}.
\]
\end{IEEEproof}

\begin{IEEEproof}[Proof of Proposition~\ref{optimalPolicy.inverse.property}]
It is easy to see that $\eta_s(x)=\sum_{i=1}^N \bar{\kappa}_s^{(i-1)}(x)$ for all $N\ge M_s(x)$. So $\eta_s$ is continuous, strictly increasing, and convex (Proposition~\ref{conjugate.property}).

Observing that $\bar{\kappa}_s^{(i)}(x)=\kappa_s^{(i)}(x)$ for all $x\ge 0$ and $i<M_s(x)$ and that $\bar{\kappa}_s^{(M_s(x))}(x)=\kappa_s^{(M_s(x))}(x)=0$ for $x$ satisfying
\[
M_s(x)
= \frac{\ln(r'(0)/r'(x))}{\ln s},
\]
we immediately have
\[
\eta_s(x)
= \sum_{i=1}^{M_s(x)} \kappa_{s^{i-1}}(x)
= \sum_{i=1}^{\tilde{M}_s(x)} \kappa_{s^{i-1}}(x).
\]
\end{IEEEproof}

\section{Important Lemmas}

\begin{lemma}\label{concavityIteration}
For a policy $\sigma\in\Sigma$ and a real-valued function $g$ on $[0,c]$, we define the function
\[
h(x)
\eqdef r(\sigma(x))+g(\flip{\sigma}(x)).
\]
If $\sigma$ is normal, $g$ is nondecreasing, Lipschitz, and concave on $[0,c]$, and $g'(\flip{\sigma}(x))\le r'(\sigma(x))$ almost everywhere, then $h$ is nondecreasing, Lipschitz, and concave on $[0,c]$, and $h'(x)\le r'(\sigma(x))$ almost everywhere.
\end{lemma}

\begin{IEEEproof}
Since $h$ is nondecreasing and Lipschitz on $[0,c]$ (Proposition~\ref{normalPolicy.property}),  it is absolutely continuous and hence differentiable a.e.\ \cite[Lemma~6.1.3 and Cor.~6.1.5]{heil_introduction_2019}, and so are $\sigma$, $r$, and $g$.
Thus, the derivative of $h$ can be computed by the differentiation rules, in particular, the chain rule \cite[Th.~6.5.2]{heil_introduction_2019}.
Specifically, we have
\begin{eqnarray*}
h'(x)
&= &r'(\sigma(x))\sigma'(x)+g'(\flip{\sigma}(x))\flip{\sigma}'(x)\\
&= &r'(\sigma(x))+\flip{\sigma}'(x)(g'(\flip{\sigma}(x))-r'(\sigma(x))) \quad\text{a.e.},
\end{eqnarray*}
which implies $h'(x)\le r'(\sigma(x))$ a.e. because $\flip{\sigma}'$ is nonnegative a.e.\ (Proposition~\ref{normalPolicy.property}).

Let $A$ be the common set on which $h'(x)$ exists and $g'(\flip{\sigma}(x))\le r'(\sigma(x))$ holds true.
It is clear that $A$ is measurable and its Lebesgue measure is $c$.
Note that $\sigma'$ and $\flip{\sigma}'$ are both nonnegative on $A$, and $\sigma'(x)$, $r'(\sigma(x))$, and $g'(\flip{\sigma}(x))$ are all nonincreasing on $A$ (Proposition~\ref{normalPolicy.property} and \cite[Th.~1.4]{gruber_convex_2007}).
For any $x,y\in A$ such that $x<y$, we have
\begin{eqnarray*}
h'(y)-h'(x)
&= &r'(\sigma(y))\sigma'(y)+g'(\flip{\sigma}(y))\flip{\sigma}'(y)\\
& &- r'(\sigma(x))\sigma'(x)-g'(\flip{\sigma}(x))\flip{\sigma}'(x)\\
&= &(r'(\sigma(y))-r'(\sigma(x)))\sigma'(y)\\
& &+ r'(\sigma(x))(\sigma'(y)-\sigma'(x))\\
& &+ (g'(\flip{\sigma}(y))-g'(\flip{\sigma}(x)))\flip{\sigma}'(y)\\
& &+ g'(\flip{\sigma}(x))(\flip{\sigma}'(y)-\flip{\sigma}'(x))\\
&= &(r'(\sigma(y))-r'(\sigma(x)))\sigma'(y)\\
& &+ (g'(\flip{\sigma}(y))-g'(\flip{\sigma}(x)))\flip{\sigma}'(y)\\
& &+ (\sigma'(y)-\sigma'(x))(r'(\sigma(x))-g'(\flip{\sigma}(x)))\\
&\le &0,
\end{eqnarray*}
which implies that $h'$ is nonincreasing on $A$.
Therefore, $h$ is concave on $[0,c]$ (Proposition~\ref{acf.convexity}).
\end{IEEEproof}

\begin{lemma}[{\cite[Appx.~C]{shaviv_universally_2016} and \cite[Eq.~(9)]{arafa_online_2018}}]\label{bernoulli.performance}
The asymptotic expected average reward of a stationary policy $\sigma\in\Sigma$ with respect to the Bernoulli energy arrival distribution $\bernoulli_p$ is
\[
T_\sigma(c)
= \sum_{i=1}^\infty p(1-p)^{i-1} r(\sigma(\flip{\sigma}^{(i-1)}(c))).\label{bernoulli.asymptotic.reward}
\]
\end{lemma}

\begin{lemma}\label{bernoulli.performance.derivative}
$T_\omega'(c)=pr'(\omega(c))$.
\end{lemma}

\begin{IEEEproof}
By Lemma~\ref{bernoulli.performance} and Theorem~\ref{optimalPolicy.property},
\[
T_\sigma(c)
\eqdef \sum_{i=1}^{M(\omega(c))} p(1-p)^{i-1} r(\omega(\flip{\omega}^{(i-1)}(c))).
\]
We have
\[
T_\sigma'(c)
= \sum_{i=1}^{M(\omega(c))} p(1-p)^{i-1} r'(\omega(\flip{\omega}^{(i-1)}(c))) \frac{\diff \omega(\flip{\omega}^{(i-1)}(c))}{\diff c}
\]
for almost every $c\ge 0$.
Since
\begin{eqnarray*}[rclqTl]
\multicolumn{3}{l}{(1-p)^{i-1} r'(\omega(\flip{\omega}^{(i-1)}(c)))}\\
\quad &= &(1-p)^{i-1} r'(\kappa_{1/(1-p)^{i-1}}(\omega(c))) &(Theorem~\ref{optimalPolicy.property})\\
&= &r'(\omega(c))
\end{eqnarray*}
for $1\le i\le M(\omega(c))$, it follows that
\begin{eqnarray*}
T_\sigma'(c)
&= &\sum_{i=1}^{M(\omega(c))} p r'(\omega(c)) \frac{\diff \omega(\flip{\omega}^{(i-1)}(c))}{\diff c}\\
&= &p r'(\omega(c)) \frac{\diff \sum_{i=1}^{M(\omega(c))} \omega(\flip{\omega}^{(i-1)}(c))}{\diff c}\\
&= &pr'(\omega(c))\frac{\diff c}{\diff c}
= pr'(\omega(c))
\end{eqnarray*}
for almost every $c\ge 0$.
\end{IEEEproof}

\section{Useful Facts}

\begin{proposition}\label{acf.convexity}
If $f$ is an absolutely continuous real-valued function on a closed interval $I=[a,b]$ and $f'$ is nondecreasing (resp., nonincreasing) a.e.\ on the set of points where it exists, then $f$ is convex (resp., concave) on $I$.
\end{proposition}

\begin{IEEEproof}
Since $f$ is absolutely continuous, it is differentiable a.e.\ and satisfies
\[
f(x)-f(a)
= \int_a^x f'(s)\diff s\quad \text{(\cite[Th.~6.4.2]{heil_introduction_2019})}.
\]
Then for any $a\le x<y\le b$ and any $t\in (0,1)$,
\begin{eqnarray*}
\multicolumn{3}{l}{(1-t)f(x)+tf(y)-f(z)}\\
\qquad &= &(1-t)(f(x)-f(z))+t(f(y)-f(z))\\
&= &-(1-t)\int_x^z f'(s)\diff s+t\int_z^y f'(s)\diff s\\
&= &-(1-t)\int_{[x,z]\cap A} f'(s)\diff s+t\int_{[z,y]\cap A} f'(s)\diff s\\
&\ge &-(1-t)\int_{[x,z]\cap A} g_z\diff s+t\int_{[z,y]\cap A} g_z\diff s\\
&= &g_z[-(1-t)(z-x)+t(y-z)]\\
&= &g_z[(1-t)x+ty-z]
= 0,
\end{eqnarray*}
where $z=(1-t)x+ty$, $g_z=\sup_{s\in [x,z]\cap A} f'(s)$, and $A$ is the set of points where $f'$ exists and is nondecreasing.
Therefore, $f$ is convex on $I$.
\end{IEEEproof}

\begin{proposition}\label{convexFunction.unbounded}
Let $f$ be a nonnegative and strictly decreasing function on $\nnreal$ (so $f(+\infty) = \lim_{x\to +\infty} f(x)$ is well defined).
If $f^{-1}(sf(x))$, defined on $[f^{-1}(f(0)/s),+\infty)$, is convex for some $s\in (1,f(0)/f(+\infty))$, then $f(+\infty) = 0$.
\end{proposition}

\begin{IEEEproof}
Let $g(x)=f^{-1}(sf(x))$ and choose an arbitrary $x_1\in [f^{-1}(f(0)/s),+\infty)$.
If $f(+\infty)>0$, then
\[
g(x)
\le f^{-1}(sf(+\infty))
< +\infty,
\]
that is, $g$ is bounded.
On the other hand, since $g(x)$ is convex,
\[
g(x_1)
= g\left(\frac{x_1}{x}x\right)
\le \frac{x_1}{x} g(x)
\]
for all $x\ge x_1$. So
\[
g(x)
\ge \frac{g(x_1)}{x_1} x,
\]
which implies that $g$ is unbounded, a contradiction to the assumption.
Therefore, $f(+\infty)=0$.
\end{IEEEproof}

%

\begin{proposition}\label{convexityOfInverse}
Let $f$ be a strictly increasing real-valued functions on some convex subset of $\real$.
Then $f$ is convex iff $f^{-1}$ is concave.
\end{proposition}

\begin{IEEEproof}
Since $f$ is strictly increasing, we have
\[
\{(x,y)\colon y\ge f(x)\}
= \{(x,y)\colon x\le f^{-1}(y)\},
\]
that is, the epigraph of $y=f(x)$ is exactly the hypograph of $x=f^{-1}(y)$.
Therefore $f$ is convex iff $f^{-1}$ is concave.
\end{IEEEproof}

\bibliographystyle{IEEEtran}

\begin{thebibliography}{00}
	
	\bibitem{SMJG10}
	V.~Sharma, U.~Mukherji, V.~Joseph, and S.~Gupta, ``Optimal
	energy management policies for energy harvesting sensor nodes," {\em
		IEEE Trans.~Wireless Commun.}, vol.~9, no.~4, pp.~1326--1336,
	Apr.~2010.
	

	
	\bibitem{OTYUY11}
	O.~Ozel, K.~Tutuncuoglu, J.~Yang, S.~Ulukus, and A.~Yener, ``Transmission
	with energy harvesting nodes in fading wireless channels: Optimal
	policies," {\em IEEE J.~Sel.~Areas Commun.}, vol.~29, no.~8, pp.~1732--1743,
	Sep.~2011.
	
	\bibitem{YU12}
	J.~Yang and S.~Ulukus, ``Optimal packet scheduling in an energy harvesting
	communication system," {\em IEEE Trans.~Commun.}, vol.~60, no.~1, pp.~220--230, Jan.~2012.
	
	\bibitem{TY12}
	K.~Tutuncuoglu and A.~Yener, ``Optimum transmission policies for battery
	limited energy harvesting nodes," {\em IEEE Trans.~Wireless Commun.},
	vol.~11, no.~3, pp.~1180--1189, Mar.~2012.
	
		
	
	\bibitem{HZ12}
	C.~K.~Ho and R.~Zhang, ``Optimal energy allocation for wireless
	communications with energy harvesting constraints," {\em IEEE Trans.~Signal
		Process.}, vol.~60, no.~9, pp.~4808--4818, Sep.~2012.
	

	
	\bibitem{OU12}	
	O.~Ozel and S.~Ulukus, ``Achieving {AWGN} capacity under stochastic energy harvesting," {\em IEEE Trans. Inf. Theory}, vol.~58, no.~10, pp.~6471--6483, Oct.~2012. 
	
	
	
	
	\bibitem{BGD13}
	P.~Blasco, D.~Gunduz, and M.~Dohler, ``A learning theoretic approach
	to energy harvesting communication system optimization," {\em IEEE Trans.~Wireless Commun.}, vol.~12, no.~4, pp.~1872--1882, Apr.~2013.
	
	
	
	\bibitem{WL13}
	Q.~Wang and M.~Liu, ``When simplicity meets optimality: Efficient
	transmission power control with stochastic energy harvesting," in {\em Proc.
		IEEE INFOCOM}, Turin, Italy, Apr.~14 - 19, 2013, pp.~580--584.
	
		
	
	\bibitem{SK13}
	R.~Srivastava and C.~E.~Koksal, ``Basic performance limits and tradeoffs
	in energy-harvesting sensor nodes with finite data and energy storage,"
	{\em IEEE/ACM Trans.~Netw.}, vol.~21, no.~4, pp.~1049--1062, Aug.~2013.
	
		
	
	\bibitem{XZ14}
	J.~Xu and R.~Zhang, ``Throughput optimal policies for energy harvesting
	wireless transmitters with non-ideal circuit power," {\em IEEE J.~Sel.~Areas
		Commun.}, vol.~32, no.~2, pp.~322--332, Feb.~2014.
	

	
	\bibitem{RSV14}
	R.~Rajesh, V.~Sharma, and P.~Viswanath, ``Capacity of Gaussian channels with energy harvesting and processing cost," {\em IEEE Trans. Inf. Theory}, vol.~60, no.~5, pp.~2563--2575, May 2014.
	
	
	
	\bibitem{UYESZGH15}	
	S.~Ulukus, A.~Yener, E.~Erkip, O.~Simeone, M.~Zorzi, P.~Grover, and K.~Huang, ``Energy harvesting wireless communications: A review of recent advances," {\em IEEE J.~Sel.~Areas
		Commun.}, vol.~33, no.~3, pp.~360--381, Mar.~2015.
	
	
	
	\bibitem{DFO15}
	Y.~Dong, F.~Farnia, and A.~\"{O}zg\"{u}r, ``Near optimal energy control and
	approximate capacity of energy harvesting communication," {\em IEEE J.~Sel.~Areas Commun.}, vol.~33, no.~3, pp.~540--557, Mar.~2015.
	
	

	
	\bibitem{shaviv_universally_2016}
	D.~Shaviv and A.~\"{O}zg\"{u}r, ``Universally near optimal
	online power control for energy harvesting nodes,''
	\emph{IEEE J.~Sel.~Areas Commun.}, vol.~34, no.~12, pp. 3620--3631, Dec. 2016.
	
	\bibitem{arafa_online_2018}
	A.~Arafa, A.~Baknina, and S.~Ulukus, ``Online fixed
	fraction policies in energy harvesting communication systems,''
	\emph{IEEE Trans.~Wireless Commun.},
	vol.~17, no.~5, pp. 2975--2986, May 2018.
	
	
	
	








	 











\bibitem{ZC19}
A.~Zibaeenejad and J.~Chen, ``The optimal power control policy for an energy harvesting system with look-ahead: Bernoulli energy arrivals,"  in {\em Proc.  IEEE  Int.  Symp.  Inform. 	Theory (ISIT)},   Paris, France, Jul. 7 - 12, 2019, pp.~116--120.

	\bibitem{WZJC19}
Y.~Wang, A.~Zibaeenejad, Y.~Jing, and J.~Chen, ``On the optimality of the greedy policy for battery
limited energy harvesting communications," arXiv:1909.07895.


















\bibitem{boyd_convex_2004}
S.~P.~Boyd and L.~Vandenberghe, \emph{Convex Optimization}.\hskip 1em plus 0.5em minus 0.4em\relax 
Cambridge, UK; New York: Cambridge University Press, 2004.

\bibitem{gruber_convex_2007}
P.~M. Gruber, \emph{Convex and Discrete Geometry}, ser. Grundlehren der
mathematischen {Wissenschaften}.\hskip 1em plus 0.5em minus 0.4em\relax
Berlin; New York: Springer, 2007, no. v. 336.

\bibitem{puterman_markov_2005}
M.~L.~Puterman, \emph{Markov Decision Processes: Discrete Stochastic Dynamic Programming}.\hskip 1em plus 0.5em minus 0.4em\relax
Hoboken, NJ: Wiley-Interscience, 2005.

\bibitem{yang_ehpart2_2019}
S.~Yang and J.~Chen, ``Universally near optimal online power control policies for battery
limited energy harvesting communications: Performance analysis and
comparison,'' in preparation.

\bibitem{kurano_existence_1989}
M.~Kurano, ``The existence of a minimum pair of state and policy for Mrkov decision processes under the hypothesis of Doeblin," {\em SIAM Journal on Control and Optimization}, vol.~27, no.~2, pp.~296--307, Mar.~1989.


\bibitem{heil_introduction_2019}
C.~Heil, \emph{Introduction to Real Analysis}, ser. Graduate Texts in Mathematics.\hskip 1em plus 0.5em minus 0.4em\relax
Cham: Springer International Publishing, 2019, vol.~280.

	
\end{thebibliography}

\end{document}